\documentclass[draft]{agujournal2019}
\usepackage{url} 
\usepackage{lineno}
\usepackage{amsmath}
\usepackage[inline]{trackchanges} 
\usepackage{soul}
\usepackage{float}
\draftfalse

\journalname{Geophysical Research Letters}

\begin{document}


\title{Butterfly Effect and the Kinetic Energy Cascade in Probabilistic Machine Learning Weather Prediction Models}

%
%
\authors{Jiakai Chen\affil{1}, Joel Oskarsson\affil{2}, Simon Driscoll\affil{3}, and Sebastian Schemm\affil{3}}

\affiliation{1}{Department of Physics, University of Cambridge, Cambridge, UK}
\affiliation{2}{ETH AI Center, ETH Zurich, Zurich, Switzerland}
\affiliation{3}{Department of Applied Mathematics and Theoretical Physics, University of Cambridge, Cambridge, UK}


\correspondingauthor{Jiakai Chen}{jc2405@cantab.ac.uk}

\begin{keypoints}
\item Purely machine-learning-based models produce realistic kinetic energy spectra magnitudes but do not exhibit expected upscale energy transfer
\item The stochastic ensemble configuration of the hybrid NeuralGCM model reproduces upscale kinetic energy transfer but encoder-level noise injection underestimates mesoscale energy
\item Machine-learning-based weather models exhibit upscale error growth but fail to reproduce the rapid initial ensemble-spread growth of small-scale errors

\end{keypoints}
%
%

\begin{abstract}

This study analyses kinetic energy (KE) spectra, difference kinetic energy (DKE) spectra, and signatures of KE transfer across spatial scales in four state-of-the-art probabilistic machine learning weather prediction (MLWP) models—NeuralGCM-ENS, FourCastNet 3, AIFS-ENS, and GenCast. Results are compared with those from the physics-based numerical weather prediction model IFS-ENS. While NeuralGCM-ENS successfully reproduces the expected upscale transfer of KE, noise injection at its encoder stage underestimates mesoscale KE. Conversely, AIFS-ENS, GenCast, and FourCastNet 3 produce realistic KE spectral magnitudes but do not capture the expected upscale transfer of KE. In particular, AIFS-ENS and GenCast, which employ spatially uncorrelated stochastic perturbations, exhibit enhanced accumulation of KE at high wavenumbers. All examined models exhibit upscale error growth, reflected by the progressive shift of the DKE spectral peak toward larger wavelengths over time. However, the MLWP models struggle to reproduce the rapid initial growth of ensemble spread at small spatial scales associated with the butterfly effect. The results show that MLWP models can misrepresent the known scale transfer of kinetic energy despite producing skilful weather forecasts.
\end{abstract}

\section*{Plain Language Summary}
This paper investigates whether cutting-edge AI weather models reproduce established principles of atmospheric fluid dynamics. We tested four recent AI weather models to see if they can reproduce the ``butterfly effect"—where tiny errors grow over time to cause large changes in weather patterns—and how they transfer energy between different length scales. We found that while AI models correctly show errors growing to larger scales over time, they struggle to capture the rapid initial growth of very small errors. Additionally, while some purely data-driven models produce realistic amounts of total energy, they fail to show realistic transfer of energy between length scales. This suggests that future AI weather models require better physical constraints to reliably predict weather patterns and extreme events.

%
%

\section{Introduction}
Reliable weather forecasting supports early warnings and planning across sectors such as transport, energy, and agriculture. Traditional numerical weather prediction (NWP) solves the equations governing atmospheric evolution but remains computationally intensive despite advances in computing, data assimilation, and parameterisation \cite{int1,nwp1,nwp2,nwp2,int3}.In recent years, data-driven methods for weather forecasting have emerged as a promising new direction \cite{int4, int5}. These models train deep neural networks on historical atmospheric datasets such as ERA5 \cite{era5} to predict atmospheric evolution from an initial condition. MLWP models’ out-of-sample forecasts, usually measured by metrics such as root-mean-square error (RMSE) and Continuous Ranked Probability Score (CRPS), have been shown to outperform predictions from state-of-the-art NWP models for up to 10 days \cite{pangu, graphcast, wb2, bench, aurora, fengwu} also at high resolution. Crucially, once trained, these data-driven models operate up to $10^4$ to $10^5$ times faster than modern NWP models \cite{pangu, graphcast}. 

Most early MLWP models and studies of their physical consistency focus on deterministic forecasting, in which a single prediction is produced from a given initial atmospheric state. These models are generally trained without explicit physical or dynamical constraints, and the physical consistency of their forecasts remains uncertain. For example, it was observed that many MLWP models like Pangu-Weather \cite{pangu} and Aurora \cite{aurora} were not able to produce the rapid growth in ensemble variance expected from the butterfly effect \cite{lorenz} when their inputs are given tiny perturbations \cite{selz1}. It was also found that MLWP models struggle in representing dynamical balance relationships of geostrophic flows \cite{ke_fcn_gc}. Furthermore, evaluations of deterministic MLWP models showed that, despite accurately capturing the locations of weather systems, their forecasts were overly smooth and tended to underestimate extreme wind and precipitation intensities \cite{simon}. 

Unlike deterministic models, recent probabilistic MLWP models such as GenCast \cite{gencast}, FourCastNet 3 \cite{fcn3}, and stochastic NeuralGCM (hereafter NeuralGCM-ENS) \cite{ngcm} generate ensembles of atmospheric trajectories. Traditional ensemble forecasting perturbs both the initial conditions and the parameters governing subgrid-scale processes in NWP models \cite{ensemble,ensemble1,ensemble2,ensemble3}, whereas some probabilistic MLWP models learn distributions over possible trajectories and generate members from a common initial condition through internal stochastic sampling \cite{fcn3}.
These probabilistic models often employ fundamentally different architectures from deterministic models. For example, GenCast implements a diffusion-based model, and has been shown to outperform earlier deterministic models in terms of producing higher-resolution features and more physically realistic power spectra \cite{gencast}. This paper evaluates the physical consistency of recent probabilistic MLWP weather models, focusing on the butterfly effect and multi-scale kinetic energy transfer. 

\section{Characteristics of the Kinetic Energy and Difference Kinetic Energy Spectrum}

\subsection{Butterfly Effect and DKE}

The butterfly effect describes the amplification of small initial perturbations into substantial forecast differences \cite{lorenz}. Globally averaged difference kinetic energy (DKE), defined here as the sum of the ensemble variances of the zonal and meridional winds, measures this ensemble spread (for a more formal definition see Section\,3.2). Weak perturbations initially grow rapidly through convection before spreading through gravity waves and geostrophic adjustment and subsequently growing upscale at synoptic scales \cite{whydke1,whydke2,selz2,initial2}. Their initial growth rate increases as the perturbation magnitude decreases \cite{initial1}.
However, deterministic MLWP models such as Pangu-Weather \cite{pangu} have been found not to reproduce this rapid initial growth \cite{selz2,selz1}.

In the spherical harmonic representation of DKE, the butterfly effect appears
as a shift of the spectral peak toward larger scales as small-scale errors
saturate, a process termed ``upscale error growth'' \cite{upscale1,upscale2,upscale3}. Numerical experiments show that, when initialized from slightly perturbed initial states, the peak of the DKE spectrum produced by deterministic AI models such as Pangu-Weather \cite{pangu} and Aurora \cite{aurora} exhibits shifts inconsistent with the butterfly effect \cite{selz1, selz2}. 

These studies primarily evaluate deterministic MLWP models by constructing ensembles initialized with prescribed perturbations. The choice of initial perturbations can affect ensemble spread and probabilistic forecast skill. For Pangu-Weather, random-field perturbations designed to preserve linear balances yielded lower CRPS than Gaussian-noise and IFS-derived perturbations, although all three approaches underestimate forecast uncertainty \cite{noise1}. 

This study focuses on recent probabilistic models, which are designed and trained to generate ensembles whose spread represents the range of plausible forecast outcomes and are therefore expected to produce more realistic ensemble spread than manually perturbed deterministic models. A recent paper evaluated the ability of GenCast, a probabilistic model, to reproduce the butterfly effect through analysis of its DKE spectra \cite{kim}, suggesting that GenCast underestimates the growth rates of DKE spectrum towards the extremes of very large or very small length scales. Our study extends existing work by comparing multiple probabilistic MLWP models that differ in both their architectures—including transformers and graph neural networks—and their probabilistic formulations, such as diffusion-based generation and noise injection with CRPS-based training.

\subsection{Upscale energy cascade and kinetic energy spectrum}

Rapid rotation and stable stratification make free-atmospheric flow behave approximately as two-dimensional turbulence, except near the planetary boundary layer \cite{2d,2d1,3d}. Two-dimensional turbulence exhibits a net upscale transfer of energy from smaller to larger length scales \cite{uet1,uet2,uet3,uet4,uet5}, a phenomenon referred to as upscale energy cascade. It was also observed experimentally \cite{expt1, expt2} and numerically \cite{num1, num2} that at long wavelengths, 2D turbulence follows the Kolmogorov-Kraichnan scaling \cite{Kraichnan, dim_analysis} where $E(n) \sim n^{-5/3}$ for wavenumber $n$, while in the limit of short wavelengths $E(n)$ follows a steeper scaling law of $E(n)\sim n^{-3}$. 
In the atmosphere, the transition occurs near $n\sim100$, corresponding to approximately 400\,km\,\cite{atm_tran}.
Recent numerical experiments observed that many deterministic MLWP models reproduce the expected KE spectra at large scales but underestimate small-scale KE \cite{ke_pangu,ke_fcn_gc}.

While most existing studies have focused on deterministic models, recent probabilistic models employ distinct architectures and stochastic formulations that may produce more realistic KE spectra. For example, NeuralGCM-ENS injects noise at every time step of the atmospheric state's evolution by seeding its machine-learned physics module with new random fields, which could shift the KE spectra of the predicted atmospheric states. To date, it remains open to what extend the new probabilistic models reproduce the cross-scale energy transfer and the butterfly effect.

\section{Experimental design}

\subsection{Data}

All ML weather models in this study were initialised from the ERA5 reanalysis dataset \cite{era5} at 00:00 UTC on 26 June 2021 and run forward for five days. This initialisation time was chosen because it coincided with a period of strong convective activity over the North American continent\cite{selz1}. For each MLWP model, we analysed a 50-member ensemble, matching the ensemble size of IFS-ENS. Before analysis, we conservatively remapped the forecast fields from all models onto a common N360 Gaussian grid with 360 longitudes and 180 latitudes (equivalent to a 1° global resolution) using the \textit{pyshtools} library, thereby preserving the global means of the kinetic energy and difference kinetic energy fields.

\subsection{Calculations of Kinetic Energy and Difference Kinetic Energy}
\label{KE and DKE}

We refer to the kinetic energy per unit mass at each grid point as simply kinetic energy (KE), and the sum of variances across ensemble members of the east-west wind (u) and north-south wind (v) at each grid point the difference kinetic energy (DKE), both in units of $\mathrm{m^2s^{-2}}$:

\begin{align}
\mathrm{KE}(\mathbf{r},\tau) &= \frac{1}{2}\left(u^2 + v^2\right)  \\
\mathrm{DKE}(\mathbf{r},\tau) &= \operatorname{var}(u) + \operatorname{var}(v)
\end{align}

In all subsequent analyses, we examine velocity fields on the 2D surface at the 500 hPa pressure level. This level has been extensively studied in meteorological research because it lies in the mid-troposphere, where large-scale atmospheric flow strongly influences and steers surface weather systems \cite{steering1, steering2, steering3, steering4}. 

We calculate the KE spectra $\mathrm{KE}(n, \tau)$ for a given wavenumber $n$ and lead time $\tau$ following the methodology of NCAR Technical Note NCAR/TN-388+STR \cite{ncar}, which we outline briefly below.

Let $\delta(\mathbf{r},\tau)$ be the divergence and $\zeta(\mathbf{r},\tau)$ the magnitude of curl of the velocity field at each grid point. Using the Python library \textit{pyshtools} \cite{pyshtools}, we then calculate the spherical harmonic coefficients $\zeta_n^m$ and $\delta_n^m$ for the decomposition into spherical harmonic modes $ Y_n^m(\lambda, \theta)$ of degree $n$ and order $m$. It can then be shown \cite{ncar} that the KE spectra at wavenumber $n$, ${KE}_n$, is given by:

\begin{equation}
\label{ke_eqn}
\mathrm{KE}(n,\tau) = \frac{a^2}{4n(n+1)} \Bigg[
\zeta_n^0 (\zeta_n^0)^* + \delta_n^0 (\delta_n^0)^*
+ 2 \sum_{m=1}^n \zeta_n^m (\zeta_n^m)^*
+ 2 \sum_{m=1}^n \delta_n^m (\delta_n^m)^*
\Bigg]
\end{equation}
with $a$ being the average radius of Earth, and asterisks denoting complex conjugation. The decomposition of DKE, denoted by $\mathrm{DKE}(n,\tau)$, into spectral components is done in an identical manner. Following conservative remapping to the N360 grid, we compute the spectra over degrees $1\leq n\leq179$, where $n=179$ is the Nyquist limit imposed by the 360-point longitudinal grid.

\subsection{Models}
\label{Models_section}

\subsubsection{IFS-ENS}
IFS-ENS is ECMWF's 50-member NWP ensemble, initialized using the Ensemble of Data Assimilation \cite{ecmwf,eda}. We obtained the IFS-ENS forecasts from the WeatherBench 2 platform \cite{wb2}, which sourced them from the TIGGE archive \cite{tigge}. The forecasts have a horizontal resolution of $0.25^\circ$ and are provided at 6-hour intervals.

\subsubsection{NeuralGCM-ENS}
NeuralGCM is a hybrid model combining a physics-based dynamical core for large-scale flow with learned representations of subgrid processes \cite{ngcm}.
NeuralGCM-ENS (the ensemble version of NeuralGCM) was trained using a combination of grid-point and spectral CRPS losses, and generates an ensemble of forecasts from a single, unperturbed initial condition. The spectral component was restricted to wavenumbers up to 80 because higher wavenumbers are filtered for stability in its dynamical core \cite{ngcm}. Stochasticity is introduced through Gaussian random fields with learned spatial and temporal correlations, which are supplied as additional inputs to both the encoder and the learned physics module. We ran NeuralGCM-ENS to obtain a 50-member ensemble, matching the size of IFS-ENS ensemble. The highest resolution available for NeuralGCM is $1.4^\circ$, with outputs at 1-hour intervals. 

\subsubsection{FourCastNet 3}
FourCastNet 3 (FCN3) is a purely data-driven probabilistic AI model \cite{fcn3}.  
Building on the spectral-loss approach previously used in NeuralGCM, FCN3 combines spatial CRPS with spectral CRPS computed from spherical harmonic coefficients. Unlike NeuralGCM, FCN3 applies the spectral loss across all resolved frequencies and variables, encouraging realistic spatial correlations and power spectra \cite{ngcm,fcn3}. 
FCN3 introduces stochasticity through a latent field with spatial and temporal
correlations generated by spherical diffusion.
We ran FCN3 at a resolution of $0.25^\circ$, with forecast fields produced at 6-hour intervals.

\subsubsection{GenCast}
GenCast is a data-driven, transformer-based diffusion model that generates
atmospheric states from spatially uncorrelated Gaussian noise, conditioned on
the two preceding states \cite{gencast}.
GenCast was reported to produce more realistic power spectra than an ensemble generated by perturbing other MLWP models like GraphCast \cite{graphcast}, addressing the common problem of excessive smoothing and blurring of atmospheric features. We downloaded GenCast forecast data from Google DeepMind's WeatherNext platform \cite{wndataset}. The forecasts have a horizontal resolution of $0.25^\circ$ and are provided at 12-hour intervals.

\subsubsection{AIFS-ENS}

AIFS-ENS \cite{aifsens} is the probabilistic version of ECMWF's Artificial Intelligence Forecasting System, a transformer-based weather forecasting model trained to produce ensemble forecasts using a loss function based on the Continuous Ranked Probability Score (CRPS). Training with CRPS effectively addressed blurring as it removed the need to collapse uncertain outcomes into a single mean state \cite{aifsens}. For each ensemble member, AIFS-ENS samples spatially uncorrelated Gaussian noise on the latent grid used by its transformer processor. We use version 1.0 of AIFS-ENS, which gives output at a resolution of $0.25^\circ$ and 6-hour intervals.

\section{Results}
\subsection{Difference Kinetic Energy}

\subsubsection{Spectral DKE growth}

\begin{figure}[!htb]
\centering
\includegraphics[width=\linewidth]{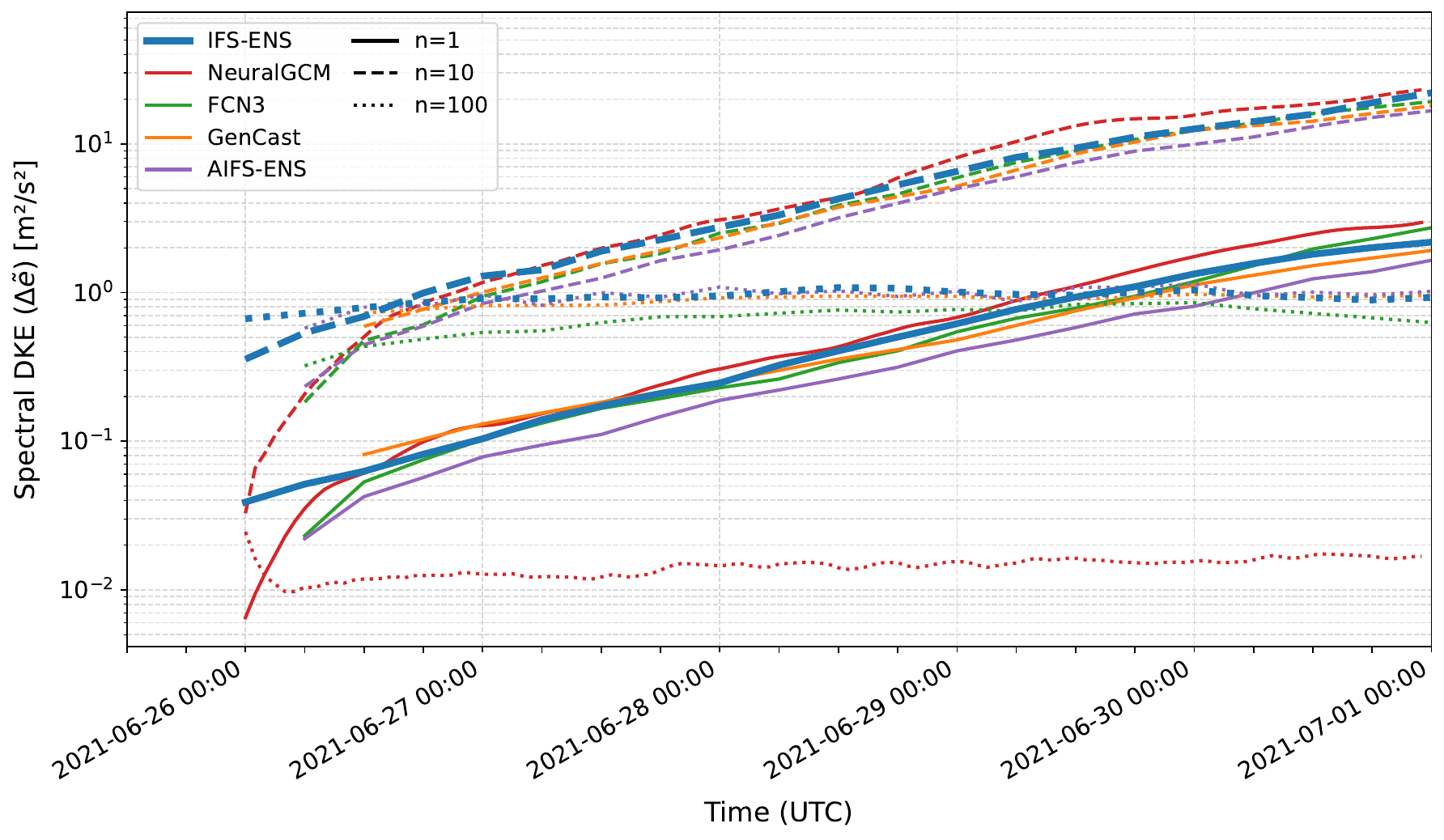}
\caption{Time evolution of spectral DKE at different spherical harmonic wavenumbers. Solid, dashed, and dotted lines represent $n=1$ ($\lambda \approx 40000$ km), $n=10$ ($\lambda \approx 4000$ km), and $n=100$ ($\lambda \approx 400$ km), respectively.}
\label{DKE_spec_K}
\end{figure}

Fig.\,\ref{DKE_spec_K} shows the growth of DKE at spherical harmonic wavenumbers of $n=1$ (solid lines), $n=10$ (dashed lines) and $n=100$ (dotted lines) separately. Studies using high-resolution NWP models show that weak initial perturbations first undergo rapid amplification through convective processes at small scales \cite{lorenz2,upscale2,selz2}. The resulting errors spread to larger scales through divergent motions and gravity-wave propagation, before being transferred to balanced flow through geostrophic adjustment and subsequently growing at synoptic scales \cite{initial2, zhang2007}. Weakly perturbed ensembles therefore exhibit rapid initial error growth and eventually approach the saturation levels of strongly perturbed ensembles. By contrast, strong initial perturbations immediately introduce substantial synoptic-scale differences, which dominate their DKE from the beginning.

In Fig.\,\ref{DKE_spec_K}, we see that for $n=1$ ($\lambda \approx 40000$ km) and $n=10$ ($\lambda \approx 4000$ km), NeuralGCM-ENS, AIFS-ENS and FCN3 which have lower DKE than IFS-ENS at the first forecasted time step showed a fast initial growth in DKE as expected, converging with IFS-ENS after around 12 hours, and grew at the same rate afterwards. GenCast has a lower initial DKE level comparable to IFS-ENS from the first time step, and showed similar growth rate as the other models. Hence ensemble spread growths realistically at large length scales for all three models. 

On the other hand, at higher wavenumbers the models show a different behaviour. At $n=100$ ($\lambda \approx 400$ km), the initial level of DKE provided by ECMWF's Ensemble Data Assimilation (EDA) \cite{eda} in IFS-ENS is already saturated hence we see the DKE of IFS-ENS staying constant in time, which matches results from previous studies using other high resolution NWP models initiated with EDA ensemble \cite{selz2}. AIFS-ENS, FCN3 and GenCast have a similar initial DKE level as IFS-ENS, hence we expect their spectrum to show saturation as well, which was observed in Fig.\,\ref{DKE_spec_K}. However, while NeuralGCM-ENS has an initial DKE more than an order of magnitude weaker than IFS-ENS, it did not show the expected fast growth to saturation. Similar behaviour was observed in previous studies when the deterministic MLWP model Pangu-Weather was initialized with low DKE \cite{selz1, selz2}. This behaviour of NeuralGCM-ENS may be due to its transition of using numerical solver for large scale processes to machine-learning parameters for sub-grid processes at these high wavenumbers. The resolution of NeuralGCM-ENS is $1.4^\circ$ \cite{ngcm}, corresponding to a wavenumber of $n\sim 130$, which may explain its loss in physical consistency and lack of DKE growth as we approach that length scale. Another possible explanation is the spectral treatment used in NeuralGCM-ENS: spherical harmonic modes above wavenumber 80 were excluded from its spectral CRPS loss because these modes are filtered for stability in the dynamical core \cite{ngcm}. A plot of the total DKE integrated over all wavenumbers is given in Fig.\,S1.

\begin{figure}[!htb]
\centering
\includegraphics[width=\linewidth]{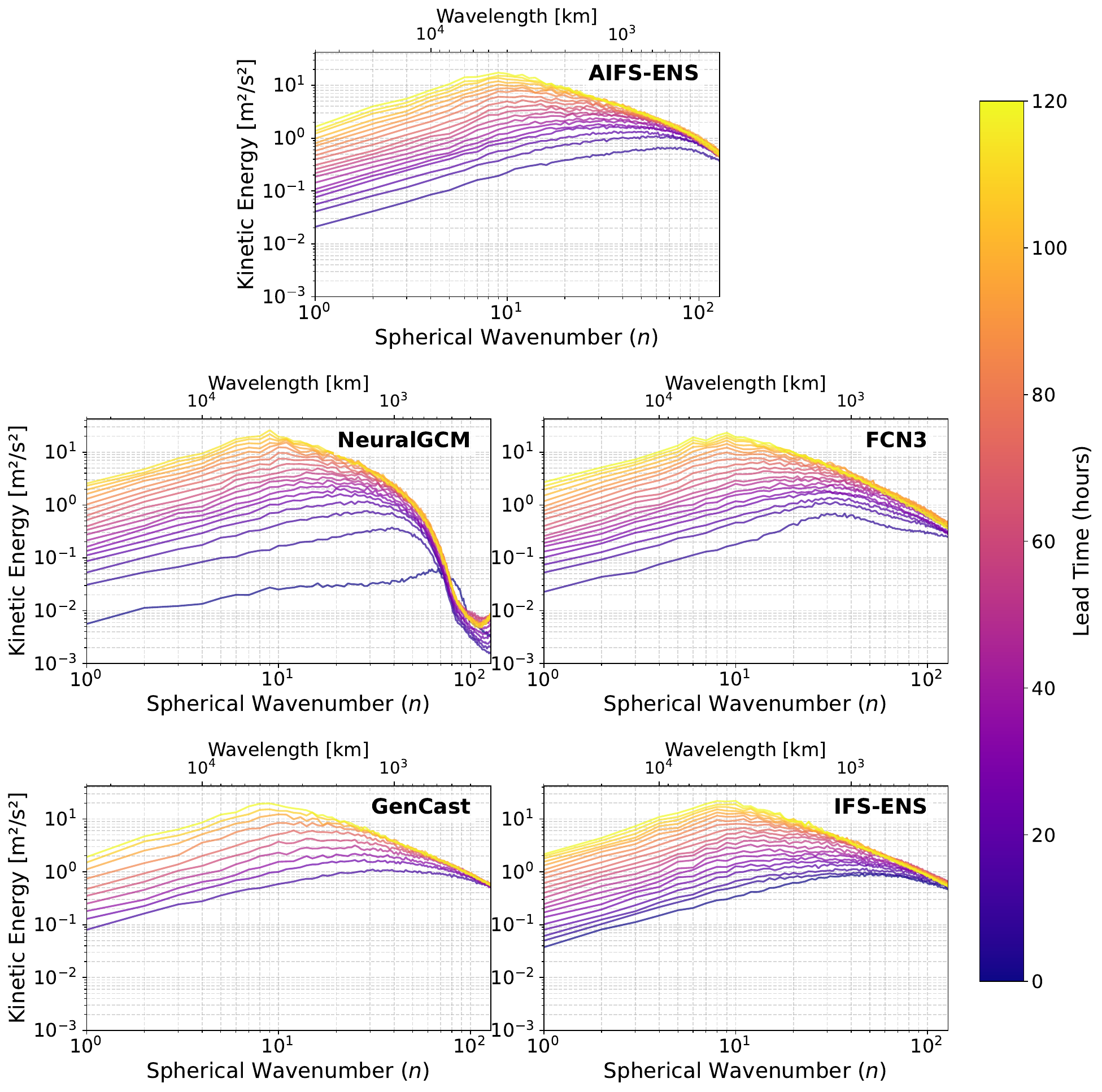}
\caption{DKE spectra at different forecast lead times.}
\label{dke_spatial}
\end{figure}

\subsubsection{Upscale error growth}
\label{upscale_error_growth_section}
Upscale error growth, which represents the growth of ensemble spread and forecast uncertainty, is evident for all four models in Fig.\,\ref{dke_spatial}, where the DKE spectral peak gradually shifts towards lower wavenumbers over time. At small spatial scales, however, NeuralGCM-ENS produces neither the expected increase in high-wavenumber DKE nor the expected DKE spectral slope. Instead, DKE decreases during the first few hours before saturating at a lower level (red dotted line in Fig.~1). In particular, NeuralGCM-ENS's spectra at the first time step show an almost flat spectrum for $n>100$, which is indicative of white noise. This suggests that the encoder-level noise injection, which provides the initial source of ensemble spread at $\tau = 0$, introduces a white-noise-like spectrum at high wavenumbers rather than perturbations consistent with balanced atmospheric dynamics. In Fig.\,S2 we show that the divergent and rotational KE spectrum of NeuralGCM-ENS merge at around $n>100$, further supporting that white noise is produced at high wavenumbers.

\subsection{Kinetic Energy Spectra}

\subsubsection{Transfer of KE}
\begin{figure}[!htb]
\centering
    \includegraphics[width = \linewidth]{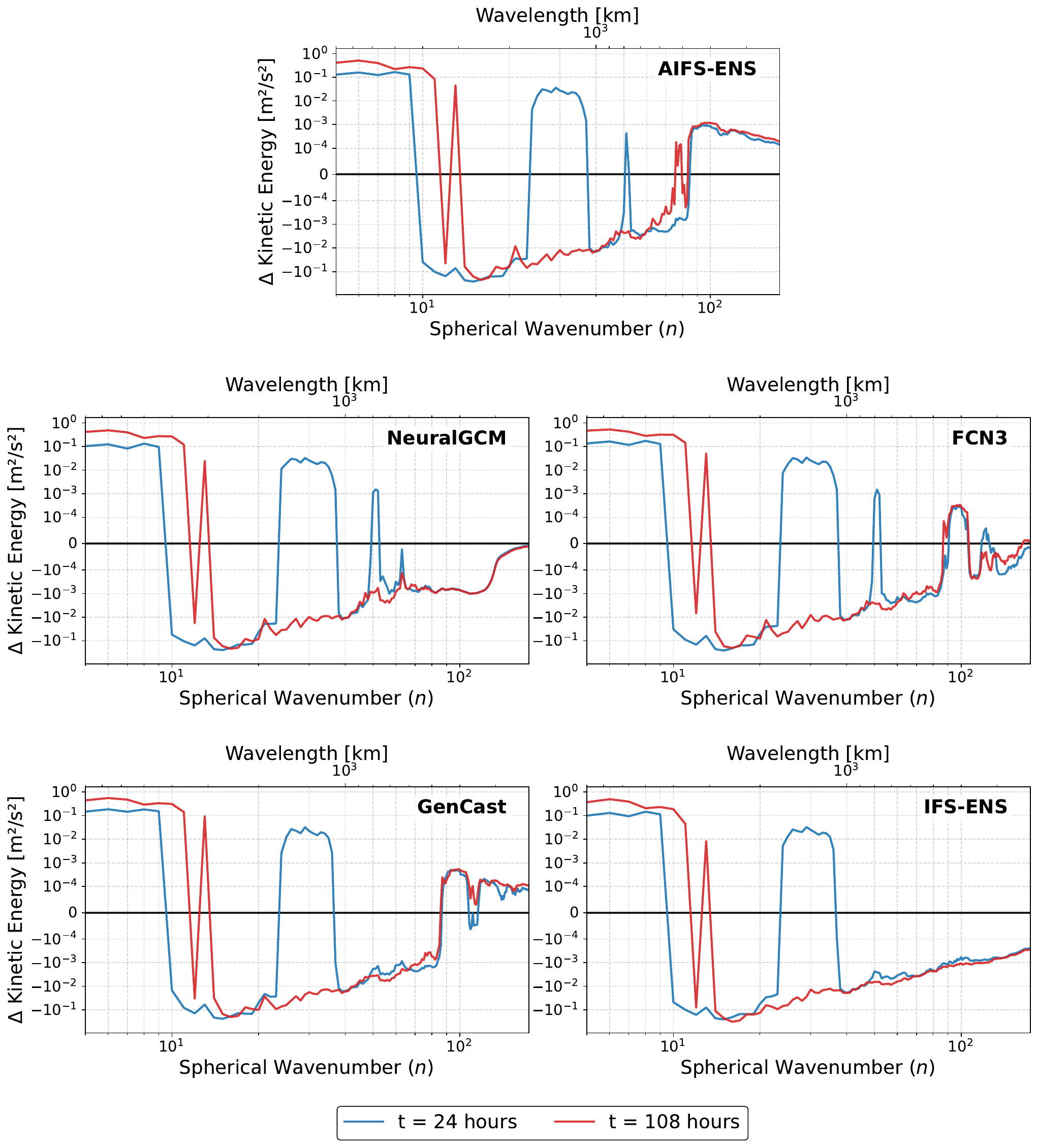}
\caption{Change of spectral KE with respect to spectra at $\tau = 0$} 
\label{KE_time_all_bin}
\end{figure}

To investigate whether each model reproduces the upscale energy cascade, we plot $\Delta\mathrm{KE}(n,\tau)$, the change of spectral KE with respect to the spectra at $\tau = 0$. For example, wavenumbers with negative $\Delta\mathrm{KE}(n,\tau)$ shows a loss in KE from that length scale with respect to initial conditions. Fig.~\ref{KE_time_all_bin} shows the plots for $\tau = 24$ hours and $\tau = 108$ hours. $\Delta\mathrm{KE}(n,\tau)$ was calculated for integer values of $n$ and subsequently smoothed using a moving average over ten nearest wavenumbers. The corresponding unsmoothed spectrum is provided in Fig.\, S5.

Initial conditions used in ensemble prediction incorporates noise due to uncertainties in observations such as instrument errors \cite{eda}, hence are largely uncorrelated between different points and corresponds to energy at short length scales. The subplot for IFS-ENS shows the expected upscale energy transfer, where for high wavenumbers we see a consistently negative $\Delta$KE at $\tau = 108$ hours except at very large scales. 

For all four MLWP models, the KE transport matches IFS-ENS closely at low wavenumbers below approximately $n=50$ but constrast substantially at higher wavenumbers. At very large spatial scales all models display positive values (below wavenumber $n\sim8$ at $\tau = 24$ hours), which increase in both height and width as lead time progresses to $\tau = 108$ hours, indicating the transfer of KE towards larger length scales.

At synoptic scales, there is a narrow plateau of positive values around  $n\sim 30$ at $\tau = 24$ hours, which is observed in all models including IFS-ENS. It is later absent at forecast lead time of $\tau = 108$ hours and it reflects the growth of atmospheric features at that characteristic scale, such as synoptic weather systems, which typically have length scales of order 1000--4000 km \cite{jetstreak}

At the smallest spatial scales the behaviour differs between IFS-ENS and  all tested models. For example, NeuralGCM-ENS reproduces the expected negative tail at high wavenumbers, which indicates KE transport towards larger scles, although with a lower magnitude at the extreme small scales and with a different slope. This is to be contrasted with the graphs of FCN3 and GenCast. FCN3 exhibits a KE spectrum at small scales ($\approx$ at wavenumbers n $\ge$ 100) that fluctuates around zero, indicating that KE is not transported properly to larger scales and with an artificial peak at n $=$ 10$^{2}$. This suggest KE accumulation at the meso-scales.
The positive tails of GenCast and AIFS-ENS suggest that the models may be generating unrealistic meso and macro-scale features and kinetic energy contained at these scales. This could be due to the use of spatially uncorrelated noise in the two models, in contrast to FCN3 and NeuralGCM-ENS. GenCast uses a new sample of isotropic Gaussian white noise to generate predictions at each forecast step through its diffusion model. AIFS-ENS samples an independent Gaussian white-noise field over each grid point, using the spatially uncorrelated noise field as latent variable inputs into its transformer.  FCN3 incorporates stochasticity through a latent random field that captures spatio-temporal correlations. This is done using a spherical diffusion process, in which independent Gaussian random coefficients are combined with spherical harmonic basis functions to generate latent fields with fixed spatial and temporal correlation length scales. This approach appears to generate a more realistic KE slope in the spectrum compared to the other tested purely data-driven models. The hybrid NeuralGCM-ENS also constructs its latent Gaussian random fields in a spherical harmonic basis, with learned spatial and temporal correlation length scales and this approach generates no KE accumulation at small scales but still an unrealistic spectral KE slope. In Fig.\, S3 we integrated the KE spectrum for $n>70$ to highlight the increasing mesoscale KE accumulated in GenCast and AIFS-ENS, and the decreasing trend in NeuralGCM-ENS and IFS-ENS.

This result suggests that NeuralGCM-ENS facilitates a somewhat more realistic interaction and upscale transfer of energy between different length scales but the spectral slope does not agree with expectations from basic fluid dynamics. This could be due to the hybrid approach taken by NeuralGCM-ENS, where the neural network is only used to parametrise sub-grid processes like cloud formation.

\subsubsection{Magnitude and scaling of KE spectra}

\begin{figure}[!htb]
\centering
    \includegraphics[width = \linewidth]{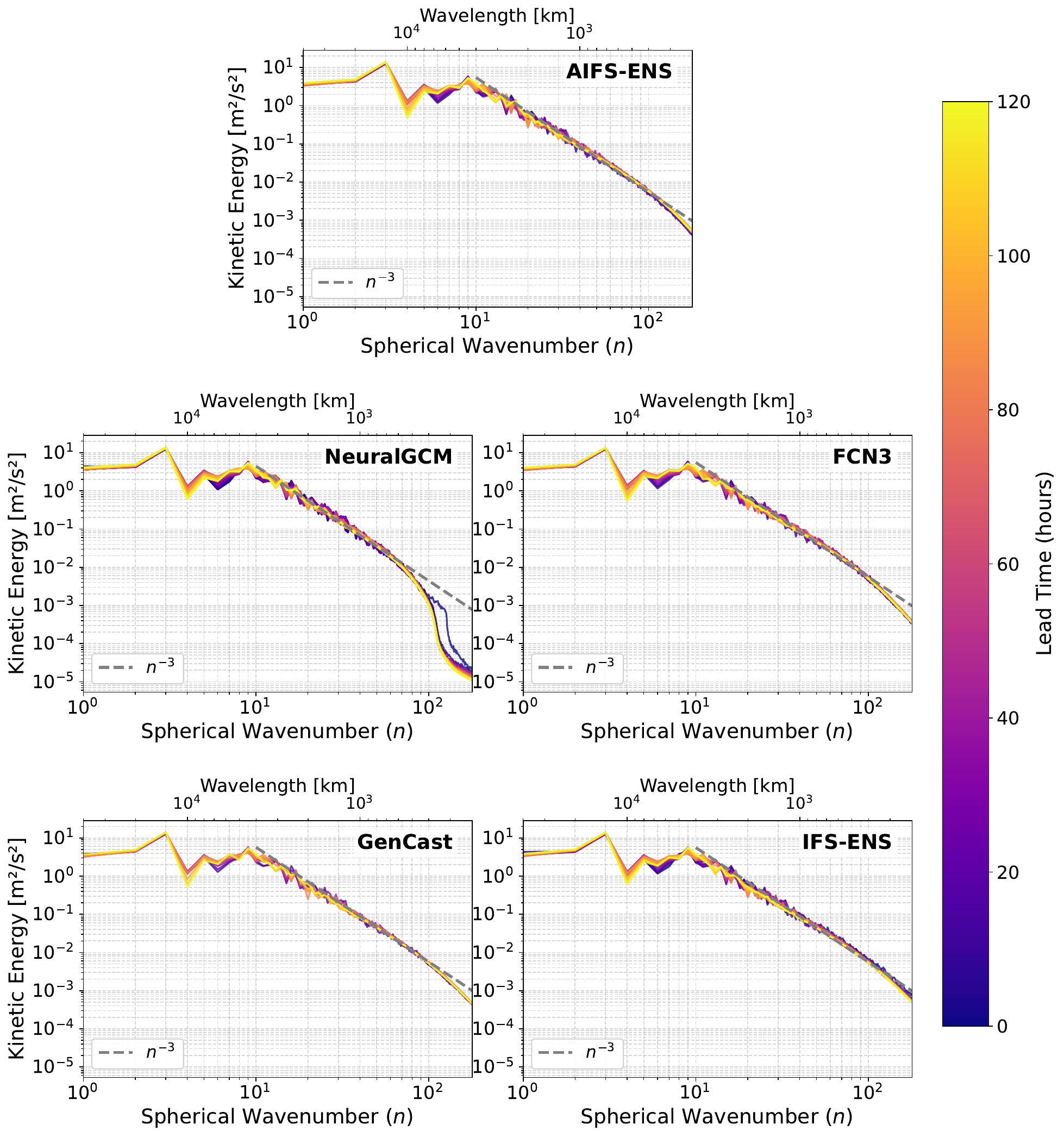}
\caption{Globally integrated KE spectrum} 
\label{ke_all}
\end{figure}

In Fig.~\ref{ke_all}, we plotted the globally integrated KE spectrum at each time for the four models. We see that the data driven models FCN3, AIFS-ENS and GenCast closely follow the expected $E(n)\sim n^{-3}$ scaling at high wavenumbers, which is also observed in IFS-ENS. For FCN3 and GenCast, this agreement is consistent with previous evaluations showing that they produce sharp forecasts with realistic power spectra, associated with their spectral-CRPS and diffusion-based formulations, respectively \cite{fcn3,gencast}. Although AIFS-ENS does not use an explicit spectral loss, previous evaluations found that its spectra remained stable with lead time and did not exhibit the progressive loss of high-wavenumber energy observed in MSE-trained models\cite{aifsens}.

By contrast, NeuralGCM-ENS follows the $n^{-3}$ scaling only up to approximately $n=70$ and underestimates KE at higher wavenumbers. This behaviour is likely related to its spectral treatment during training: the spectral CRPS loss was restricted to wavenumbers up to 80 because higher-wavenumber modes are filtered for stability in its dynamical core \cite{ngcm}. The close correspondence between the onset of KE underestimation and this cutoff suggests that the filtering may contribute to the observed loss of high-wavenumber energy. Figure~S4 provides a side-by-side comparison of the KE spectra of all models at $\tau=108$ hours.

Overall, our results suggest that although the data-driven models FCN3, AIFS-ENS and GenCast produced more realistic KE magnitudes, they were unable to reproduce realistic transfers of energy across spatial scales. In contrast, NeuralGCM-ENS exhibited upscale energy transfer, but underestimated the KE spectrum at high wavenumbers.

\section{Conclusions}
This study investigates the physical consistency of state-of-the-art probabilistic MLWP weather models through the analysis of (a) kinetic energy and (b) difference kinetic energy spectra with a focus on the upscale growth of energy, kinetic energy spectra shape and evolution and change in kinetic energy as function of wave number and forecast time. 

Concerning the upscale energy growth, all MLWP models examined, namely NeuralGCM-ENS, FCN3, AIFS-ENS and GenCast, exhibit upscale error growth in their DKE spectra (Fig.~\ref{dke_spatial}). This suggests that the models are broadly able to reproduce large-scale atmospheric dynamics qualitatively consistent with the butterfly effect. However, at small spatial scales the hybrid NeuralGCM is a notable exception. Its DKE spectrum displays a nearly flat, white-noise-like structure at initial time steps, suggesting that the injection of noise at the encoder level produced perturbations inconsistent with balanced atmospheric dynamics. There is also an unnatural transition at the interface between the machine-learned and resolved scales. Indeed, the KE spectrum (Fig.~\ref{ke_all}) indicates that NeuralGCM-ENS underestimates the magnitude of small scale KE, likely due to the machine-learned parametrization of sub-grid processes, the noise injection, which generates a flat KE spectrum at very small scales, and the cut off in the spectral loss, while producing a realistic KE spectrum at the resolved scales. 

Considering the transport of KE across scales, it is the NeuralGCM-ENS hybrid model that reproduces the expected upscale transfer of KE most closely, showing a reduction in KE at high wavenumbers over time that is similar to that seen in the IFS-ENS model, but not with expected slope in spectral space (Fig.~\ref{KE_time_all_bin}). This suggests that incorporating a physics-based dynamical core improves the physical consistency of multiscale energy interactions but predominantly at the resolved scales.
While the purely data-driven models FCN3, AIFS-ENS and GenCast produce DKE spectra with more realistic DKE at high wavenumbers (small scales), which is likely due to FCN3's spectral loss function and GenCast's diffusion-based generative architecture, they nevertheless all three fail to produce the expected upscale transfer of KE at small scale. In particular, AIFS-ENS and GenCast accumulate KE at high wavenumbers, which may be due to their use of spatially uncorrelated noise. 

Overall, our results suggest that although MLWP models achieve competitive scores in metrics such as RMSE, these metrics do not fully capture basic KE behaviour expected from observations. The DKE spectrum appears realistic in data-driven models but the KE transport across scales is not, with the hybrid model being closest to a realistic cross-scale KE behvior.  These findings raise broader questions: To what extent and how much physical consistency should machine learning weather forecasting models adhere to, acknowledging that these are not built as general purpose fluid dynamics solvers but only for the purpose of weather forecasting? To what extent does an inaccurate representation of known energy transfer across spatial scales undermine the trustworthiness of these models, given that their forecasts may nevertheless be judged to be of high quality according to other metrics?

%
%

\section*{Open Research Section}
Data for the IFS-ENS forecasts were obtained from the TIGGE archive \cite{tigge}. GenCast ensemble predictions are available via the WeatherNext dataset provided by Google Earth Engine \cite{wndataset}. 
Forecasts were generated using the official code and model checkpoints for
AIFS-ENS v1.0 (\url{https://huggingface.co/ecmwf/aifs-ens-1.0}),
FourCastNet 3 (\url{https://huggingface.co/nvidia/fourcastnet3}), and
NeuralGCM-ENS (\url{https://github.com/neuralgcm/neuralgcm}).

\section*{Conflict of Interest declaration}
The authors declare there are no conflicts of interest for this manuscript.

\acknowledgments

JO was supported by the ETH AI Center through an ETH AI Center postdoctoral fellowship. SD would like to acknowledge support from both Schmidt Sciences, LLC, and the Institute of Computing for Climate Science, University of Cambridge. JC performed most of the analysis during a master thesis project at the University of Cambridge supervised by SS and SD.


\end{document}